\documentclass[superscriptaddress,showpacs,twocolumn,preprintnumbers]{revtex4}
\begin{document}

\preprint{DPNU-02-14}

\title{Exact Formulas and Simple CP dependence of Neutrino \\
Oscillation Probabilities in Matter with Constant Density}
\author{Keiichi Kimura}
\email[E-mail address: ]{kimukei@eken.phys.nagoya-u.ac.jp}
\affiliation{Department of Physics, Nagoya University,
Nagoya, 464-8602, Japan}
\author{Akira Takamura}
\email[E-mail address: ]{takamura@eken.phys.nagoya-u.ac.jp}
\affiliation{Department of Physics, Nagoya University,
Nagoya, 464-8602, Japan}
\affiliation{Department of Mathematics, Toyota National Collage of Technology 
Eisei-cho 2-1, Toyota-shi, 471-8525, Japan}
\author{Hidekazu Yokomakura}
\email[E-mail address: ]{yoko@eken.phys.nagoya-u.ac.jp}
\affiliation{Department of Physics, Nagoya University,
Nagoya, 464-8602, Japan}


%
%

\begin{abstract}
We investigate neutrino oscillations in constant matter within the 
context of the standard three neutrino scenario.
We derive an exact and simple formula for the oscillation probability 
applicable to all channels.
In the standard parametrization,
the probability for $\nu_e$ $\to$ $\nu_{\mu}$ transition  
can be written in the form 
$P(\nu_e \to \nu_{\mu})=A_{e\mu}\cos\delta+B_{e\mu}\sin\delta+C_{e\mu}$ 
without any approximation using CP phase $\delta$.
For $\nu_{\mu}$ $\to$ $\nu_{\tau}$ transition, 
the linear term of $\cos 2\delta$ is added
and the probability can be written in the form 
$P(\nu_{\mu} \to \nu_{\tau})=A_{\mu\tau}\cos\delta+B_{\mu\tau}
\sin\delta+C_{\mu\tau}+D_{\mu\tau}\cos 2\delta$.
We give the CP dependences of the probability for other channels. 
We show that the probability for each channel in matter
has the same form with respect to $\delta$ as in vacuum.
It means that matter effects just modify the coefficients 
$A$, $B$, $C$ and $D$.
We also give the exact expression of the coefficients for each channel.
Furthermore, we show that our results with respect to CP dependences 
are reproduced from the effective mixing angles and the effective CP 
phase calculated by Zaglauer and Schwarzer. 
Through the calculation, 
a new identity is obtained by dividing the Naumov-Harrison-Scott 
identity by the Toshev identity. 

\end{abstract}

\pacs{14.60.Pq}

\maketitle

\section{Introduction}
\label{sec:intro}

Strong evidences for neutrino oscillations have been accumulated 
and they indicate the finite masses and mixings of neutrinos.  
The first evidence is the observation of zenith-angle 
dependence of the atmospheric neutrino deficit 
\cite{atmospheric}, which is consistent with  
$\nu_{\mu} \to \nu_{\tau}$ transition 
with the mass difference and the mixing as 
\begin{eqnarray}
\Delta_{32}\sim 3\times 10^{-3}{\rm eV}^2, 
\hspace{0.5cm} \sin^2 2\theta_{23}\sim 1,\label{-1}
\end{eqnarray}
where $\Delta_{ij}=m_i^2-m_j^2$ and $m_i$ is 
neutrino mass.
The second evidence is the solar neutrino deficit 
\cite{solar}, which is consistent with 
$\nu_e \to \nu_{\mu}/\nu_{\tau}$ transition.
Recent reports for the measurements of CC and NC interactions 
in the SNO experiment \cite{SNO}  
are consistent with the Standard Solar Model 
and strongly suggest the LMA MSW solution 
\begin{eqnarray}
\Delta_{21}\sim 7\times 10^{-5}{\rm eV}^2, 
\hspace{0.5cm} \sin^2 2\theta_{12}\sim 0.8 \label{0}
\end{eqnarray}
to the solar neutrino problem.
Thus, the information on neutrino masses and mixings
(Maki-Nakagawa-Sakata matrix) \cite{MNS} has been 
increased considerably.
The main physics goals in future experiments are the determination 
of the unknown parameters $\theta_{13}$ 
(only upper bound $\sin^2 2\theta_{13} < 0.1$ 
is obtained \cite{CHOOZ}) and the CP phase $\delta$ 
in addition to the high precision measurement of the parameters 
in (\ref{-1}) and (\ref{0}). 
In particular, the observation of $\delta$ is quite 
interesting from the point of view that $\delta$ is related to
the origin of the matter in the universe.
In order to get at the CP phase $\delta$,
a considerable number of studies have been 
made on CP violation \cite{CP, Cervera, Freund, Sato, Minakata2}. 
Moreover several experiments using neutrino beam 
are planned \cite{JHF, nu-factory}.
In such experiments with high energy neutrino beam, 
matter effects cannot be neglected and prevent 
the precise measurement of pure CP effects.
In this paper we derive an exact and simple formula of 
oscillation probabilities for all channels. 

Before giving our results, let us give the present status of 
exact formulas 
associated with oscillation probability in constant matter. 
The mass eigenvalues have been given by Barger {\it et al.}
 \cite{Barger1}.
Zaglauer and Schwarzer \cite{Zaglauer} have calculated  
effective mixing angles and effective CP phase 
in the standard parametrization.
Furthermore, the elements of the MNS matrix have been calculated 
by Xing \cite{Xing} in parametrization independent way.
However, these exact formulas obtained by diagonalizing 
the Hamiltonian, are complicated and 
it is not easy to obtain useful information on 
matter effects and the CP phase.
Ohlsson and Snellman \cite{Ohlsson} have directly 
calculated the amplitude without diagonalizing the 
Hamiltonian although the CP phase is not considered 
(which is similar to our result as shown in Sec.\ref{sec:3}). 

Recently, by comparing the Hamiltonian 
in matter and that in vacuum, 
Naumov \cite{Naumov} and Harrison and Scott \cite{Harrison} 
have presented simple and elegant identity 
\begin{eqnarray}
\tilde{\Delta}_{12}\tilde{\Delta}_{23}\tilde{\Delta}_{31}
\tilde{J}=\Delta_{12}\Delta_{23}\Delta_{31}J,
\end{eqnarray}
where $J$ is the Jarlskog factor \cite{Jarlskog} 
and tilde represents the quantities in matter.
We have investigated the matter enhancement of $\tilde{J}$ 
in Ref. \cite{Yokomakura} by using this identity.
Furthermore, the enhancement of T-violating quantity 
$\Delta P_T=P(\nu_e \to \nu_{\mu})-P(\nu_{\mu} \to \nu_e)$ 
has been also investigated by Parke and Weiler \cite{Parke} 
and Akhmedov {\it et al.} \cite{Akhmedov}.

In our previous paper \cite{Kimura}, 
we have found other identities on the product of 
the MNS matrix elements by comparing the Hamiltonian 
in matter and that in vacuum.
From these identities, we have presented the exact formula for 
$P(\nu_e \to \nu_{\mu})$ and we have definitely separated 
pure CP effects from matter effects.
In the standard parametrization, we have also shown 
that the probability for $\nu_e$ $\to$ $\nu_{\mu}$ transition 
can be written in the form
\footnote{It has been first pointed out by 
 Krastev and Petcov \cite{Krastev} that CP odd term is 
proportional to $\sin \delta$.
It is easy to understand from the Naumov-Harrison-Scott 
identity \cite{Naumov, Harrison}.} \cite{Kimura}
\begin{eqnarray}
P(\nu_e \to \nu_{\mu})=A_{e\mu}\cos \delta+B_{e\mu}\sin\delta+C_{e\mu}.
\end{eqnarray}
It has also been found 
that CP trajectory proposed by Minakata and Nunokawa 
\cite{Minakata} is exactly elliptic in bi-probability 
space when $\delta$ changes from $0$ to $2\pi$.

In this paper, extending our previous results \cite{Kimura}, 
we derive an exact and simple formula of  
the oscillation probabilities for all channels. 
We also find that the probability for each channel in matter 
has the same form with respect to the CP phase $\delta$ as in vacuum.
For example, the probability for $\nu_{\mu}$ $\to$ $\nu_{\tau}$
transition can be written as 
\begin{equation}
P(\nu_{\mu} \to \nu_{\tau})=A_{\mu\tau}\cos \delta+B_{\mu\tau}
\sin\delta+C_{\mu\tau}+D_{\mu\tau}\cos 2\delta,
\end{equation}
both in vacuum and in matter.
Namely, matter effects just modify 
$A_{\mu\tau}$, $B_{\mu\tau}$, $C_{\mu\tau}$ and $D_{\mu\tau}$.
We also give the exact expression of these coefficients.
Furthermore, we show that our results with respect 
to CP dependences are reproduced from the effective mixing angles 
and effective CP phase calculated by 
Zaglauer and Schwarzer \cite{Zaglauer}. 
Although the effective CP phase are complicated, 
the CP dependence of probability is simplified for all channels. 
This is guaranteed by the Toshev identity 
\begin{equation}
\tilde{s}_{23} \tilde{c}_{23} \sin \tilde{\delta}
= \tilde{s}_{23} \tilde{c}_{23} \sin \tilde{\delta}.
\end{equation}
Finally, we obtain a new identity
\begin{equation}
\tilde{\Delta}_{12}\tilde{\Delta}_{23}\tilde{\Delta}_{31}
\tilde{s}_{12}\tilde{c}_{12}\tilde{s}_{13}\tilde{c}_{13}^2
=\Delta_{12}\Delta_{23}\Delta_{31}s_{12}c_{12}s_{13}c_{13}^2,
\end{equation}
related to the 1-2 mixing and the 1-3 mixing by dividing 
the Naumov-Harrison-Scott identity by the Toshev identity. 

\section{CP dependences of Oscillation Probabilities in Vacuum}
\label{sec:2}

In this section, we review the CP dependences of the oscillation 
probabilities in vacuum.
The Hamiltonian in vacuum 
\begin{eqnarray}
H=\left(
\begin{array}{ccc}
H_{ee}  & H_{e\mu} & H_{e\tau} \\
H_{\mu e} & H_{\mu\mu} & H_{\mu\tau} \\
H_{\tau e} & H_{\tau\mu} & H_{\tau\tau}
\end{array}
\right)
\end{eqnarray}
is diagonalized as  
\begin{eqnarray}
U^{\dagger}HU
=\frac{1}{2E}\left(
\begin{array}{ccc}
0 &  &  \\
  & \Delta_{21} &  \\
  & & \Delta_{31}
\end{array}
\right), \label{0-1}
\end{eqnarray} 
where $U$ is the MNS matrix which connects 
flavor eigenstates with mass eigenstates
\footnote{Overall phase does not affect 
the probability from (\ref{2}).
By using this, the eigenvalues of the Hamiltonian 
are written with only mass differences 
subtracting $m_1^2$.}.
Then, the amplitude of $\nu_{\alpha}$ 
to $\nu_{\beta}$ transition after transporting the distance $L$ is 
\begin{eqnarray}
A(\nu_{\alpha} \to \nu_{\beta})
=\sum_{i=1}^3
U_{\alpha i}^* e^{-i\frac{\Delta_{i1}}{2E}L}U_{\beta i}, \label{1}
\end{eqnarray}
where $\alpha$ and $\beta$ denote flavor indices $e$, $\mu$ or $\tau$.
And the oscillation probability is given by 
\begin{eqnarray}
\lefteqn{P(\nu_{\alpha} \to \nu_{\beta})
=|A(\nu_{\alpha} \to \nu_{\beta})|^2} 
\nonumber \\  
&&\hspace{-0.5cm}=\delta_{\alpha\beta}
-4\sum_{(ij)}^{{\rm cyclic}}{\rm Re}J_{\alpha\beta}^{ij}
\sin^2 \Delta_{ij}^{\prime} 
\pm 2\sum_{(ij)}^{{\rm cyclic}}J
\sin 2\Delta_{ij}^{\prime} \label{2},
\end{eqnarray}
where 
\begin{eqnarray}
&&J_{\alpha\beta}^{ij}\equiv
U_{\alpha i}U_{\beta i}^*(U_{\alpha j}U_{\beta j}^*)^*,
\quad
J\equiv {\rm Im}J_{e\mu}^{12}, \\ 
&&\Delta_{ij}^{\prime}\equiv \frac{\Delta_{ij}L}{4E}
\equiv \frac{(m_i^2-m_j^2) L}{4E} \label{3}.
\end{eqnarray}
The $\pm$ sign of the third term takes $-(+)$ 
in the case that $(\alpha, \beta)$ is given by 
(anti) cyclic permutation of $(e,\mu)$.
The cyclic sum is over $(ij)=(12), (23), (31)$.

Next let us calculate ${\rm Re} J_{e\mu}^{ij}$ and $J$ 
to study the CP dependences of probabilities 
in the standard parametrization \cite{PDG},  
\begin{widetext}
\begin{eqnarray}
U_{\alpha i}=\left(
\begin{array}{ccc}
c_{12}c_{13} & s_{12}c_{13} & s_{13}e^{-i\delta} \\
-s_{12}c_{23}-c_{12}s_{23}s_{13}e^{i\delta}
& c_{12}c_{23}-s_{12}s_{23}s_{13}e^{i\delta}
& s_{23}c_{13} \\
s_{12}s_{23}-c_{12}c_{23}s_{13}e^{i\delta}
& -c_{12}s_{23}-s_{12}c_{23}s_{13}e^{i\delta}
& c_{23}c_{13}
\end{array}
\right), 
\end{eqnarray}
where $s_{ij}\equiv \sin \theta_{ij}$,
$c_{ij}\equiv \cos \theta_{ij}$.

${\rm Re} J_{e\mu}^{ij}$ are given by 
\begin{eqnarray}
{\rm Re}J_{e\mu}^{12}&=&-(c_{12}^2-s_{12}^2)J_r \cos\delta
+s_{12}^2 c_{12}^2 c_{13}^2(s_{23}^2 s_{13}^2-c_{23}^2), 
\label{4} \\
{\rm Re}J_{e\mu}^{23}&=&J_r \cos\delta
-s_{12}^2 s_{23}^2 s_{13}^2 c_{13}^2, \\
{\rm Re}J_{e\mu}^{31}&=&-J_r \cos\delta
-c_{12}^2 s_{23}^2 s_{13}^2 c_{13}^2, \label{5}
\end{eqnarray}
and ${\rm Re} J_{e\tau}^{ij}$ are given by 
\begin{eqnarray}
{\rm Re}J_{e\tau}^{12}&=&(c_{12}^2-s_{12}^2)J_r \cos\delta
+s_{12}^2 c_{12}^2 c_{13}^2(c_{23}^2 s_{13}^2-s_{23}^2), \\
{\rm Re}J_{e\tau}^{23}&=&-J_r \cos\delta
-s_{12}^2 c_{23}^2 s_{13}^2 c_{13}^2, \\
{\rm Re}J_{e\tau}^{31}&=&J_r \cos\delta
-c_{12}^2 c_{23}^2 s_{13}^2 c_{13}^2, 
\end{eqnarray}
and ${\rm Re} J_{\mu\tau}^{ij}$ are given by 
\begin{eqnarray}
{\rm Re}J_{\mu\tau}^{12}&=&s_{12}^2 c_{12}^2 s_{23}^2 c_{23}^2 
(1+s_{13}^2+s_{13}^4)-(s_{12}^2 c_{12}^2 +s_{23}^2 c_{23}^2)s_{13}^2
\nonumber \\
&-&(c_{12}^2-s_{12}^2)(c_{23}^2-s_{23}^2)
s_{12}c_{12}s_{23}c_{23}s_{13}(1+s_{13}^2)\cos\delta
+2s_{12}^2 c_{12}^2 s_{23}^2 c_{23}^2 s_{13}^2 \cos 2\delta, \\
{\rm Re}J_{\mu\tau}^{23}&=&-(c_{23}^2-s_{23}^2)J_r \cos\delta
+s_{23}^2 c_{23}^2 c_{13}^2(s_{12}^2 s_{13}^2-c_{12}^2), \\
{\rm Re}J_{\mu\tau}^{31}&=&(c_{23}^2-s_{23}^2)J_r \cos\delta
+s_{23}^2 c_{23}^2 c_{13}^2(c_{12}^2 s_{13}^2-s_{12}^2),
\end{eqnarray}
\end{widetext}
where $J_r\equiv s_{12}c_{12}s_{23}c_{23}s_{13}c_{13}^2$.
We also obtain the Jarlskog factor,  
\begin{eqnarray}
J=J_r \sin\delta \label{6}.
\end{eqnarray}
It should be noted that the replacements 
$s_{23} \to c_{23}$, $c_{23} \to -s_{23}$ in 
${\rm Re}J_{e\mu}^{ij}$ leads to ${\rm Re}J_{e\tau}^{ij}$.

In summary, the CP dependences of transition probabilities 
are obtained by substituting (\ref{4})-(\ref{6}) into (\ref{2}) as
\begin{eqnarray}
&&\hspace{-0.8cm}
P(\nu_e \to \nu_{\mu})=A_{e\mu}\cos\delta+B\sin\delta+C_{e\mu}, 
\label{7} \\
&&\hspace{-0.8cm}
P(\nu_e \to \nu_{\tau})=-A_{e\mu}\cos\delta-B\sin\delta+C_{e\tau},\\
&&\hspace{-0.8cm}
P(\nu_{\mu} \to \nu_{\tau})=A_{\mu\tau}\cos\delta
+B\sin\delta+C_{\mu\tau}+D\cos 2\delta. 
\end{eqnarray}
$P(\nu_e \to \nu_{\mu})$ and $P(\nu_e \to \nu_{\tau})$ 
have only linear terms of $\cos\delta$ and $\sin\delta$.
On the other hand, $P(\nu_{\mu} \to \nu_{\tau})$ has 
also a term proportional to $\cos 2\delta$ in addition to 
the terms of $\cos \delta$ and $\sin \delta$.

Next, let us consider survival probabilities.
One can calculate the CP dependences from the unitarity and 
the transition probabilities. 
Those are given by 
\begin{eqnarray}
P(\nu_e \to \nu_e)&=&C_{ee}, \\
P(\nu_{\mu} \to \nu_{\mu})&=&A_{\mu\mu}\cos\delta
+C_{\mu\mu}-D\cos 2\delta, \\
P(\nu_{\tau} \to \nu_{\tau})&=&A_{\tau\tau}\cos\delta
+C_{\tau\tau}-D\cos 2\delta, \label{8}
\end{eqnarray}
where the coefficients can be calculated from 
(\ref{4})-(\ref{6}).
Note that the term proportional to $\sin \delta$ 
disappears in survival probabilities.
In particular, not only $\sin \delta$ term but also 
$\cos \delta$ term does not exist in $P(\nu_e \to \nu_e)$ 
\cite{Kuo}.
In $P(\nu_{\mu} \to \nu_{\mu})$ and 
$P(\nu_{\tau} \to \nu_{\tau})$, the linear term of $\cos 2\delta$ 
has opposite sign to that in $P(\nu_{\mu} \to \nu_{\tau})$.
The probabilities for CP conjugate and T conjugate 
channels are obtained from (\ref{7})-(\ref{8}) 
by the exchange $\delta \to -\delta$.

\section{Derivation of Exact Formula  
for {\boldmath $P(\nu_{\alpha} \to \nu_{\beta})$}}
\label{sec:3}

In this section, we present three identities 
on the product of the MNS matrix elements. 
Then, we derive an exact formula for the oscillation 
probability.
Furthermore, we confirm that  
the Naumov-Harrison-Scott identity can be also derived 
from our identities.

The Hamiltonian in matter $\tilde{H}$ is given by  
\begin{eqnarray}
\tilde{H}=H+
\frac{1}{2E}\left(
\begin{array}{ccc}
a &  &  \\
  & 0 &  \\
  & & 0
\end{array}
\right)
\label{9},
\end{eqnarray}
where $H$ is the Hamiltonian in vacuum and is diagonalized as (\ref{0-1}), 
$a\equiv 2\sqrt{2}G_F N_e E$ is the matter potential, 
$G_F$ is the Fermi constant and $N_e$ is the electron density in matter.
Note that the second term in (\ref{9}) is energy independent 
because the matter potential $a$ is proportional to the energy $E$.
The Hamiltonian in matter is diagonalized by the effective MNS matrix as 
\begin{eqnarray}
\tilde{U}^{\dagger}\tilde{H}\tilde{U}
=\frac{1}{2E}\left(
\begin{array}{ccc}
\lambda_1 &  &  \\
  & \lambda_2 &  \\
  & & \lambda_3
\end{array}
\right) \label{10}, 
\end{eqnarray}
where $\lambda_i$ is mass eigenvalue in matter.
The amplitude and probability for $\nu_{\alpha} \to 
\nu_{\beta}$ transition in matter are obtained 
by the replacements $m_i^2 \to \lambda_i$ and 
$U \to \tilde{U}$ in (\ref{1})-(\ref{3}).

As one of the approaches to derive the probability, 
one can calculate single $\tilde{U}_{\alpha i}$ by directly 
diagonalizing $\tilde{H}$ and calculate $\tilde{J}_{\alpha\beta}^{ij}$, 
which is the product of four $\tilde{U}$s.
However, the expressions for the probabilities obtained in this approach are  
complicated and it is not easy to extract the 
information on matter effects and $\delta$.
Here, we derive the probability using another 
approach.
Note that $\tilde{U}_{\alpha i}\tilde{U}_{\beta i}^*$ appears 
in the amplitude in (\ref{1}) and the probability in (\ref{2}).
So, we only have to calculate the product
$\tilde{U}_{\alpha i}\tilde{U}_{\beta i}^*$ 
and the expression for single $\tilde{U}_{\alpha i}$ 
is not necessary.
We use the relation (\ref{9}) in order to calculate 
$\tilde{U}_{\alpha i}\tilde{U}_{\beta i}^*$.
We introduce $\tilde{p}_{\alpha\beta}$ and $\tilde{q}_{\alpha\beta}$
as 
\begin{eqnarray}
\tilde{p}_{\alpha\beta}&=&2E
\tilde{H}_{\alpha\beta}, \label{10-1} \\
\tilde{q}_{\alpha\beta}&=&
(2E)^2 \tilde{{\cal H}}_{\alpha\beta} 
=
(2E)^2(\tilde{H}_{\gamma\beta}\tilde{H}_{\alpha\gamma}-
\tilde{H}_{\alpha\beta}\tilde{H}_{\gamma\gamma}),
\label{10-2}
\end{eqnarray}
where $(\alpha \beta \gamma)=(e\mu\tau),(\mu\tau e),(\tau e\mu)$.
Then, we obtain 
three kinds of identities on 
$\tilde{U}_{\alpha i}\tilde{U}_{\beta i}^*$.
First, from unitarity we obtain  
\begin{eqnarray}
\tilde{U}_{\alpha 1}\tilde{U}_{\beta 1}^*
+\tilde{U}_{\alpha 2}\tilde{U}_{\beta 2}^*
+\tilde{U}_{\alpha 3}\tilde{U}_{\beta 3}^*
=\delta_{\alpha\beta} \label{13}. 
\end{eqnarray}
Second, $2E\tilde{H}_{\alpha\beta}=\tilde{p}_{\alpha\beta}$ in 
(\ref{10-1}), 
which is also seen in Ref. \cite{Xing2}, is 
rewritten as 
\begin{eqnarray}
\lambda_1 \tilde{U}_{\alpha 1}\tilde{U}_{\beta 1}^*
+\lambda_2 \tilde{U}_{\alpha 2}\tilde{U}_{\beta 2}^*
+\lambda_3 \tilde{U}_{\alpha 3}\tilde{U}_{\beta 3}^*
=\tilde{p}_{\alpha\beta}
\label{14}. 
\end{eqnarray}
Third, $(2E)^2\tilde{{\cal H}}_{\alpha\beta}=
\tilde{q}_{\alpha\beta}$ in (\ref{10-2}), is rewritten as  
\begin{eqnarray}
\lambda_2 \lambda_3 \tilde{U}_{\alpha 1}\tilde{U}_{\beta 1}^*
+\lambda_3 \lambda_1 \tilde{U}_{\alpha 2}\tilde{U}_{\beta 2}^*
+\lambda_1 \lambda_2 \tilde{U}_{\alpha 3}\tilde{U}_{\beta 3}^*
=\tilde{q}_{\alpha\beta}
\label{15},
\end{eqnarray}
where we use the relation 
\begin{eqnarray}
(2E)^2\tilde{\cal{H}}&=&(2E)^2\tilde{H}^{-1}({\rm det}\tilde{H})
\nonumber \\
&=&\tilde{U}{\rm diag}\left(\frac{1}{\lambda_1},\frac{1}{\lambda_2},
\frac{1}{\lambda_3}\right)\tilde{U}^{\dagger}\times 
\lambda_1 \lambda_2 \lambda_3.
\end{eqnarray} 

Eqs. (\ref{13})-(\ref{15}) on 
$\tilde{U}_{\alpha i}\tilde{U}_{\beta i}^*$ 
can be simply solved as 
\begin{eqnarray}
\tilde{U}_{\alpha i}\tilde{U}_{\beta i}^*=
\frac{\tilde{p}_{\alpha\beta}\lambda_i+\tilde{q}_{\alpha\beta}
-\delta_{\alpha\beta}\lambda_i(\lambda_j+\lambda_k)}
{\tilde{\Delta}_{ji}\tilde{\Delta}_{ki}}, \label{16}
\end{eqnarray}
where $(ijk)$ takes $(123), (231), (312)$.
Note that $\tilde{p}_{\alpha\beta}$ and $\tilde{q}_{\alpha\beta}$ 
become constants in the following cases,
\begin{eqnarray}
\tilde{p}_{\alpha\beta}&=&p_{\alpha\beta}\quad 
(\alpha \neq e \,\, {\rm or} \,\, \beta \neq e), \\ 
\tilde{q}_{\alpha\beta}&=&q_{\alpha\beta}\quad
(\alpha=e \,\, {\rm or} \,\, \beta=e)
\end{eqnarray}
from (\ref{9}).
In other cases, $\tilde{H}_{ee}$ is included in 
$\tilde{p}_{\alpha\beta}$ and $\tilde{q}_{\alpha\beta}$,  
and they depend on $a$ explicitly.
Similar expression for 
$\tilde{U}_{\alpha i}\tilde{U}_{\beta i}^*$ in (\ref{16})
has been obtained by Ohlsson and Snellman in Ref. \cite{Ohlsson} and  
also by Harrison and Scott \cite{Harrison2} 
although the method of derivation is different.
We have given an exact and simple 
formula of $P(\nu_e \to \nu_{\mu})$ for the first time 
in Ref. \cite{Kimura}. 
Here, we give the general formula of oscillation probability 
$P(\nu_{\alpha} \to \nu_{\beta})$ applicable to all channels 
by calculating 
${\rm Re}\tilde{J}_{\alpha\beta}^{ij}$ and $\tilde{J}$ 
from $\tilde{J}_{\alpha\beta}^{ij}
=\tilde{U}_{\alpha i}\tilde{U}_{\beta i}^*
(\tilde{U}_{\alpha j}\tilde{U}_{\beta j}^*)^*$.
In the case of $\alpha \neq \beta$, the expression is given by 
\begin{equation}
P(\nu_{\alpha} \to \nu_{\beta})
=-4\sum_{(ij)}^{{\rm cyclic}}{\rm Re}\tilde{J}_{\alpha\beta}^{ij}
\sin^2 \tilde{\Delta}_{ij}^{\prime} 
\pm 2\sum_{(ij)}^{{\rm cyclic}}\tilde{J}
\sin 2\tilde{\Delta}_{ij}^{\prime},
\label{17}
\end{equation}
where 
\begin{equation}
{\rm Re}\tilde{J}_{\alpha\beta}^{ij}=
\frac{|\tilde{p}_{\alpha\beta}|^2 \lambda_i \lambda_j
+|\tilde{q}_{\alpha\beta}|^2
+{\rm Re}(\tilde{p}_{\alpha\beta}\tilde{q}_{\alpha\beta}^*)
(\lambda_i+\lambda_j)}
{\tilde{\Delta}_{ij}\tilde{\Delta}_{12}\tilde{\Delta}_{23}
\tilde{\Delta}_{31}},  
\end{equation}
\begin{eqnarray}
\tilde{J}=\frac{{\rm Im}
(\tilde{p}_{e\mu}\tilde{q}_{e\mu}^*)}
{\tilde{\Delta}_{12}\tilde{\Delta}_{23}
\tilde{\Delta}_{31}} \label{18}, \quad
\tilde{\Delta}_{ij}^{\prime}\equiv\frac{\tilde{\Delta}_{ij}L}{4E}.
\end{eqnarray}
The concrete expression for $\lambda_i$ 
\footnote{When the calculation from amplitude to probability in matter, 
overall phase does not affect the probability.
We remove the matter effects from NC interaction in (\ref{9})
and  $m_1^2$ in (\ref{10}).
Therefore, the expression for $\lambda_i$ does not depend on 
$m_1^2$ but mass differences.}
is calculated 
by the MNS matrix elements and the masses in vacuum as 
\begin{eqnarray}
\lambda_1&=&\frac{1}{3}s-\frac{1}{3}
\sqrt{s^2-3t}\left[u+\sqrt{3(1-u^2)}\right], \\
\lambda_2&=&\frac{1}{3}s-\frac{1}{3}
\sqrt{s^2-3t}\left[u-\sqrt{3(1-u^2)}\right], \\
\lambda_3&=&\frac{1}{3}s+\frac{2}{3}
u\sqrt{s^2-3t}, 
\end{eqnarray}
where $s, t, u$ are given by 
\begin{eqnarray}
&&\hspace{-0.8cm}s=\Delta_{21}+\Delta_{31}+a , \\
&&\hspace{-0.8cm}t=\Delta_{21}\Delta_{31}
+a[\Delta_{21}(1-s_{12}^2 c_{13}^2)+\Delta_{31}(1-s_{13}^2)], \\
&&\hspace{-0.8cm}u=\cos \left[\frac{1}{3}\cos^{-1}
\left(\frac{2s^3-9st+27a\Delta_{21}\Delta_{31}c_{12}^2 c_{13}^2}
{2(s^2-3t)^{3/2}}\right)\right], 
\end{eqnarray}
in the standard parametrization \cite{Barger1, Zaglauer, Xing}.
One can see that the effective masses in matter 
do not depend on the CP phase $\delta$.

Finally, we reproduce the Naumov-Harrison-Scott identity from 
(\ref{18}) as 
\begin{eqnarray}
\tilde{\Delta}_{12}\tilde{\Delta}_{23}\tilde{\Delta}_{31}\tilde{J}
&=&\frac{1}{(2E)^3}{\rm Im}[\tilde{H}_{e\mu}
(\tilde{H}_{e\tau}\tilde{H}_{\tau\mu}
-\tilde{H}_{e\mu}\tilde{H}_{\tau\tau})^*] \nonumber \\
&=&\frac{1}{(2E)^3}{\rm Im}[\tilde{H}_{e\mu}\tilde{H}_{\mu\tau}
\tilde{H}_{\tau e}] \nonumber \\
&=&\frac{1}{(2E)^3}{\rm Im}[H_{e\mu}H_{\mu\tau}H_{\tau e}] \nonumber \\
&=&\Delta_{12}\Delta_{23}\Delta_{31}J.
\label{18-1}
\end{eqnarray}
Thus, the Naumov-Harrison-Scott identity is easily derived  
from the identities which we used.

\section{CP dependences of Oscillation Probabilities in Matter}
\label{sec:4}

In this section, we calculate the oscillation probabilities 
in all channels by using (\ref{17})-(\ref{18}) 
and investigate their CP dependences.
 
\subsection*{\boldmath $P(\nu_e \to \nu_{\mu})$}

Let us consider the CP dependence of 
$P(\nu_e \to \nu_{\mu})$ which is 
the best channel for the observation of 
CP violation.
First, we calculate 
$\tilde{p}_{e\mu}$ and $\tilde{q}_{e\mu}$ 
from (\ref{14}) and (\ref{15}). 
As $\tilde{p}_{e\mu}$ and $\tilde{q}_{e\mu}$ do not include 
$\tilde{H}_{ee}$,
they are equivalent to the constants $p_{e\mu}$ and $q_{e\mu}$, 
which are represented by the quantities in vacuum.
Therefore, we obtain  
\begin{eqnarray}
&&p_{e\mu}=\Delta_{21} U_{e2}U_{\mu 2}^*+\Delta_{31} U_{e3}U_{\mu 3}^*, \\
&&q_{e\mu}=\Delta_{31}\Delta_{21}U_{e1}U_{\mu 1}^*
\end{eqnarray}
from (\ref{14}) and (\ref{15}).
Next, we divide $p_{e\mu}$ and $q_{e\mu}$ 
into the terms including and not including $\delta$ as 
\begin{eqnarray}
p_{e\mu}=p_{e\mu}^a e^{-i\delta}+p_{e\mu}^b, \hspace{0.5cm}
q_{e\mu}=q_{e\mu}^a e^{-i\delta}+q_{e\mu}^b,
\end{eqnarray}
where $p_{e\mu}^a$, $p_{e\mu}^b$, $q_{e\mu}^a$ and 
$q_{e\mu}^b$ are real numbers given by 
\begin{eqnarray}
&&p_{e\mu}^a=(\Delta_{31}-\Delta_{21}s_{12}^2)
s_{23}s_{13}c_{13}, \label{19} \\
&&p_{e\mu}^b=\Delta_{21}s_{12}c_{12}c_{23}c_{13}, \\
&&q_{e\mu}^a=-\Delta_{31}\Delta_{21}c_{12}^2 s_{23}s_{13}c_{13}, \\
&&q_{e\mu}^b=-\Delta_{31}\Delta_{21}s_{12}c_{12}c_{23}c_{13},
\label{20}
\end{eqnarray}
in the standard parametrization.

The oscillation probability can be written in the form 
\begin{eqnarray}
P(\nu_e \to \nu_{\mu})=\tilde{A}_{e\mu} \cos \delta
+\tilde{B}\sin \delta+\tilde{C}_{e\mu} \label{21}
\end{eqnarray}
from (\ref{17})-(\ref{18}).
Note that $P(\nu_e \to \nu_{\mu})$ is exactly 
linear in $\sin \delta$ and $\cos \delta$.

Next, we give the exact expression for 
$\tilde{A}_{e\mu}$, $\tilde{B}$ and $\tilde{C}_{e\mu}$. 
We rewrite them in the product of 
the oscillation part which depends on $L$ 
and $(\tilde{A}_r)_{ij}$, $\tilde{B}_r$ and 
$(\tilde{C}_r)_{ij}$ as 
\begin{eqnarray}
\tilde{A}_{e\mu}&=&\sum_{(ij)}^{{\rm cyclic}}(\tilde{A}_r)_{ij}
\sin^2 \tilde{\Delta}_{ij}^{\prime},
\\
\tilde{B}&=&\sum_{(ij)}^{{\rm cyclic}}\tilde{B}_r
\sin 2\tilde{\Delta}_{ij}^{\prime} \label{22},
\\
\tilde{C}_{e\mu}&=&\sum_{(ij)}^{{\rm cyclic}}
(\tilde{C}_r)_{ij}
\sin^2 \tilde{\Delta}_{ij}^{\prime}.
\end{eqnarray}
$\tilde{B}$ is expressed in the form of the sum as (\ref{22}).
Under the condition $x+y+z=0$, the relation    
\begin{eqnarray}
\sin 2x+\sin 2y +\sin 2z=-4\sin x \sin y \sin z
\end{eqnarray}
holds and $\tilde{B}$ from (\ref{22}) is rewritten in the form of product as 
\begin{eqnarray}
\tilde{B}=\sum_{(ij)}^{{\rm cyclic}}\tilde{B}_r
\sin 2\tilde{\Delta}_{ij}^{\prime}=-4\tilde{B}_r
\sin \tilde{\Delta}_{12}^{\prime}
\sin \tilde{\Delta}_{23}^{\prime}
\sin \tilde{\Delta}_{31}^{\prime}.
\end{eqnarray}
$\tilde{A}_{e\mu}$ is rewritten in the same way.
Under the same condition as in deriving $\tilde{B}$, 
the relation  
\begin{eqnarray}
\sin^2 x=-(\sin x \sin y \cos z+\sin x \cos y \sin z)
\end{eqnarray}
holds and $\tilde{A}_{e\mu}$ is rewritten as  
\begin{eqnarray}
\lefteqn{\tilde{A}_{e\mu}=\sum_{(ij)}^{{\rm cyclic}}
(\tilde{A}_r)_{ij}\sin^2
\tilde{\Delta}_{ij}^{\prime}} \nonumber \\
&&\hspace{-0.5cm}=-\sum_{(ijk)}^{{\rm cyclic}}
[(\tilde{A}_r)_{jk}+(\tilde{A}_r)_{ki}]
\cos \tilde{\Delta}_{ij}^{\prime}
\sin \tilde{\Delta}_{jk}^{\prime}
\sin \tilde{\Delta}_{ki}^{\prime}.
\end{eqnarray}
Substituting (\ref{19})-(\ref{20}) into 
$p_{e\mu}$ and $q_{e\mu}$ in (\ref{17})-(\ref{18}), 
$\tilde{A}_{e\mu}$, $\tilde{B}$ and $\tilde{C}_{e\mu}$ 
are rewritten with the masses and mixings as 
\begin{widetext}
\begin{eqnarray}
\tilde{A}_{e\mu}&=&\sum_{(ijk)}^{{\rm cyclic}}\frac{-8[J_r\Delta_{21}
\Delta_{31}\lambda_k(\lambda_k-\Delta_{31})
+(\tilde{A}_{e\mu})_k]}
{\tilde{\Delta}_{jk}^2\tilde{\Delta}_{ki}^2} 
\cos \tilde{\Delta}_{ij}^{\prime}
\sin \tilde{\Delta}_{jk}^{\prime}
\sin \tilde{\Delta}_{ki}^{\prime},
\label{23} \\
\tilde{B}&=&\frac{8J_r \Delta_{12} \Delta_{23} \Delta_{31}}
{\tilde{\Delta}_{12} \tilde{\Delta}_{23}
\tilde{\Delta}_{31}}
\sin \tilde{\Delta}_{12}^{\prime}
\sin \tilde{\Delta}_{23}^{\prime}
\sin \tilde{\Delta}_{31}^{\prime},
\label{24} \\
\tilde{C}_{e\mu}&=&\sum_{(ij)}^{{\rm cyclic}}
\frac{-4[\Delta_{31}^2 s_{13}^2 s_{23}^2 c_{13}^2
\lambda_i \lambda_j
+(\tilde{C}_{e\mu})_{ij}]}
{\tilde{\Delta}_{ij} \tilde{\Delta}_{12}
\tilde{\Delta}_{23}\tilde{\Delta}_{31}} 
\sin^2 \tilde{\Delta}_{ij}^{\prime}
\label{25}.
\end{eqnarray}
\end{widetext}
See Appendix for the expression of 
$(\tilde{A}_{e\mu})_k$ and $(\tilde{C}_{e\mu})_{ij}$. 
Note that these expressions (\ref{23})-(\ref{25}) are still exact.
In the limit of small $\Delta_{21}$, 
$(\tilde{A}_{e\mu})_k$ and $(\tilde{C}_{e\mu})_{ij}$ 
are higher order in $\Delta_{21}$ and can be ignored.

Finally, we obtain the well known approximate formula by 
neglecting the smallest effective mass.
Considering the energy of neutrino 
($E \geq 10\,{\rm GeV}$) in neutrino factory experiment 
and the earth matter density 
($\rho \simeq 2.8\, {\rm g/{cm^3}}$), 
matter potential is given by 
\begin{eqnarray}
a&=&2\sqrt{2}G_F N_e E=7.56 \times 10^{-5} {\rm eV}^2 
\frac{\rho}{{\rm gcm^{-3}}} \frac{E}{{\rm GeV}} \nonumber \\
&\geq& 2.1 \times 10^{-3} {\rm eV}^2.
\end{eqnarray}
It means that the smallest effective mass $\lambda_1$ 
is almost equivalent to $\Delta_{21}$. 
Other effective masses  
$\lambda_2$ and $\lambda_3$, correspond to $a$ or $\Delta_{31}$.
Accordingly, the coefficients are approximated by   
\begin{eqnarray}
\tilde{A}_{e\mu}&\simeq&\frac{8J_r \Delta_{21}\Delta_{31}}
{a(\Delta_{31}-a)}
\cos \Delta_{31}^{\prime}
\sin a^{\prime}
\sin (\Delta_{31}-a)^{\prime},
 \\
\tilde{B}&\simeq&\frac{8J_r \Delta_{21} \Delta_{31}}
{a(\Delta_{31}-a)}
\sin \Delta_{31}^{\prime}
\sin a^{\prime}
\sin (\Delta_{31}-a)^{\prime}, \\
\tilde{C}_{e\mu}&\simeq&\frac{4\Delta_{31}^2 
s_{23}^2 s_{13}^2 c_{13}^2}
{(\Delta_{31}-a)^2}
\sin^2 (\Delta_{31}-a)^{\prime}.
\end{eqnarray}
Although the approximate formula derived here is 
in agreement with the ones seen in 
Ref. \cite{Cervera, Freund}, the derivation is rather simple.
The coefficients $\tilde{A}_{e\mu}$ and $\tilde{B}$ responsible for 
$\delta$ are $O(\Delta_{21}s_{13})$.
On the other hand, $\tilde{C}_{e\mu}$  
is $O(s_{13}^2)$.
Therefore, the terms related to $\delta$ become relatively 
large in the case of small $s_{13}$.

\subsection*{\boldmath $P(\nu_e \to \nu_{\tau})$}

In this subsection, we calculate 
$P(\nu_e \to \nu_{\tau})$ in the same way 
as $P(\nu_e \to \nu_{\mu})$.
We need to learn $\tilde{U}_{ei}\tilde{U}_{\tau i}^*$ 
for this purpose.

At first, we consider $\tilde{p}_{e\tau}$ and 
$\tilde{q}_{e\tau}$.
Since $\tilde{p}_{e\tau}$ and 
$\tilde{q}_{e\tau}$ do not include $\tilde{H}_{ee}$,
they are equivalent to the constants $p_{e\tau}$ and $q_{e\tau}$.
Therefore, we obtain
\begin{eqnarray}
&&p_{e\tau}=
\Delta_{21} U_{e2}U_{\tau 2}^*+\Delta_{31} U_{e3}U_{\tau 3}^*, \\
&&q_{e\tau}=\Delta_{31}\Delta_{21}U_{e1}U_{\tau 1}^*
\end{eqnarray}
from (\ref{14}) and (\ref{15}).
Next, we divide $p_{e\tau}$ and $q_{e\tau}$ 
into the terms including and not including $\delta$ as
\begin{eqnarray}
p_{e\tau}=p_{e\tau}^a e^{-i\delta}+p_{e\tau}^b, \hspace{0.5cm}
q_{e\tau}=q_{e\tau}^a e^{-i\delta}+q_{e\tau}^b,
\end{eqnarray}
where $p_{e\tau}^a$, $p_{e\tau}^b$, $q_{e\tau}^a$ and 
$q_{e\tau}^b$ are real numbers given by 
\begin{eqnarray}
&&p_{e\tau}^a=(\Delta_{31}-\Delta_{21}s_{12}^2)
c_{23}s_{13}c_{13}, \\
&&p_{e\tau}^b=-\Delta_{21}s_{12}c_{12}s_{23}c_{13}, \\
&&q_{e\tau}^a=-\Delta_{31}\Delta_{21}c_{12}^2 c_{23}s_{13}c_{13},
\\
&&q_{e\tau}^b=\Delta_{31}\Delta_{21}s_{12}c_{12}s_{23}c_{13},
\end{eqnarray}
in the standard parametrization.
Note that $p_{e\tau}$ and $q_{e\tau}$ are 
obtained by the replacements 
$s_{23} \to c_{23}, c_{23} \to -s_{23}$ in 
$p_{e\mu}$ and $q_{e\mu}$. 
The coefficients of 
$\cos \delta$ and $\sin\delta$ have the opposite sign 
by the replacements, from (\ref{23}) and (\ref{24}).
Therefore, $P(\nu_e \to \nu_{\tau})$ can be also written 
in the form 
\begin{eqnarray}
P(\nu_e \to \nu_{\tau})=-\tilde{A}_{e\mu}\cos \delta
-\tilde{B}\sin \delta+\tilde{C}_{e\tau},
\end{eqnarray}
where 
\begin{eqnarray}
\tilde{C}_{e\tau}=\sum_{(ij)}^{{\rm cyclic}}
\frac{-4[\Delta_{31}^2 s_{13}^2 c_{23}^2 c_{13}^2
\lambda_i \lambda_j
+(\tilde{C}_{e\tau})_{ij}]}
{\tilde{\Delta}_{ij} \tilde{\Delta}_{12}
\tilde{\Delta}_{23}\tilde{\Delta}_{31}}
\sin^2 \tilde{\Delta}_{ij}^{\prime}, 
\end{eqnarray}
and the concrete expression for $(\tilde{C}_{e\tau})_{ij}$ 
is given in Appendix.
As for $\tilde{A}_{e\mu}$ and $\tilde{B}$,  
see (\ref{23}) and (\ref{24}).
Note that this expression is also exact. 
In the limit of small $\Delta_{21}$, 
$(\tilde{C}_{e\tau})_{ij}$ is high order in $\Delta_{21}$ 
and can be neglected. 

As in the case of $P(\nu_e \to \nu_{\mu})$, 
we obtain the approximate formula as  
\begin{eqnarray}
\tilde{C}_{e\tau}\simeq\frac{4\Delta_{31}^2
c_{23}^2 s_{13}^2 c_{13}^2}
{(\Delta_{31}-a)^2}
\sin^2 (\Delta_{31}-a)^{\prime},
\end{eqnarray}
by neglecting the smallest effective mass 
$\lambda_1\simeq \Delta_{21}$.
As the result, the leading term $\tilde{C}_{e\tau}$ 
which is not related to 
$\delta$ is $O(s_{13}^2)$ 
as in the case of $P(\nu_e \to \nu_{\mu})$.
It means that CP effect becomes relatively large 
also in this channel. 

\subsection*{\boldmath $P(\nu_{\mu} \to \nu_{\tau})$}

In this subsection, we calculate 
$P(\nu_{\mu} \to \nu_{\tau})$.
The expression for the probability of this channel 
is useful to analyze the appearance experiments 
of MINOS \cite{MINOS} and CNGS \cite{CNGS}. 
It is also needed for the analysis of neutrino factory 
experiments in which $\nu_{\tau}$ is produced by the 
high energy $\nu_{\mu}$ beam.
In order to derive $P(\nu_{\mu} \to \nu_{\tau})$, 
we calculate $\tilde{U}_{\mu i}\tilde{U}_{\tau i}^*$.
Let us first consider $\tilde{p}_{\mu\tau}$ and 
$\tilde{q}_{\mu\tau}$ in 
$\tilde{U}_{\mu i}\tilde{U}_{\tau i}^*$.
Since $\tilde{q}_{\mu\tau}$ includes $H_{ee}$,
it is not constant and depends on the matter potential 
$a$ explicitly as
\begin{eqnarray}
\frac{1}{(2E)^2}\tilde{q}_{\mu \tau}&=&\tilde{{\cal H}}_{\mu\tau}
=\left\{H_{e\tau}H_{\mu e}
-\left(H_{ee}+\frac{a}{2E}\right)H_{\mu\tau}\right\} \nonumber \\
&=&\frac{1}{(2E)^2}(q_{\mu \tau}-ap_{\mu \tau}). 
\end{eqnarray}
Then, $\tilde{U}_{\mu i}\tilde{U}_{\tau i}^*$ is given by 
\begin{eqnarray}
\tilde{U}_{\mu i}\tilde{U}_{\tau i}^*=
\frac{\tilde{p}_{\mu\tau}\lambda_i+\tilde{q}_{\mu\tau}}
{\tilde{\Delta}_{ji}\tilde{\Delta}_{ki}}
=\frac{p_{\mu\tau}(\lambda_i-a)+q_{\mu\tau}}
{\tilde{\Delta}_{ji}\tilde{\Delta}_{ki}}.
\end{eqnarray}
Here, the different point between $P(\nu_e \to \nu_{\mu})$ 
and $P(\nu_e \to \nu_{\tau})$ is that the coefficient of 
$p_{\mu\tau}$ is not $\lambda_i$ but $\lambda_i-a$.
Fortunately, the matter potential $a$ does not depend on $\delta$,  
so we have only to investigate $p_{\mu\tau}$ and 
$q_{\mu\tau}$, in order to learn the CP dependence 
of the probability.
In the standard parametrization, we obtain   
\begin{widetext}
\begin{eqnarray}
p_{\mu \tau}=p_{\mu \tau}^a e^{-i\delta}+p_{\mu \tau}^b
+p_{\mu \tau}^c e^{i\delta}, \quad
q_{\mu \tau}=q_{\mu \tau}^a e^{-i\delta}+q_{\mu \tau}^b
+q_{\mu \tau}^c e^{i\delta}, 
\end{eqnarray}
where
\begin{eqnarray}
&&p_{\mu \tau}^a=-\Delta_{21}s_{12}c_{12}c_{23}^2 s_{13},
\quad
p_{\mu \tau}^b=[\Delta_{31}c_{13}^2
-\Delta_{21}(c_{12}^2-s_{12}^2 s_{13}^2)]s_{23}c_{23}, 
\quad
p_{\mu \tau}^c=\Delta_{21}s_{12}c_{12}s_{23}^2 s_{13}, \\
&&q_{\mu \tau}^a=\Delta_{31}\Delta_{21}s_{12}c_{12}c_{23}^2 s_{13},
\quad
q_{\mu \tau}^b=\Delta_{31}\Delta_{21}(-s_{12}^2+c_{12}^2 s_{13}^2)
s_{23}c_{23}, \quad
q_{\mu \tau}^c=-\Delta_{31}\Delta_{21}s_{12}c_{12}s_{23}^2 s_{13}.
\end{eqnarray}
And then, the transition probability can be written in 
the form
\begin{eqnarray}
P_{\mu\tau}
=\tilde{A}_{\mu\tau}\cos \delta
+\tilde{B}\sin \delta+\tilde{C}_{\mu\tau}
+\tilde{D}\cos 2\delta,
\end{eqnarray}
where we obtain the expression for the coefficients after 
some straightforward calculations as 
\begin{eqnarray}
\tilde{A}_{\mu\tau}&=&\sum_{(ijk)}^{{\rm cyclic}}\frac{-8[
J_r \Delta_{21}\Delta_{31}(\lambda_k-a)(\lambda_k-a-\Delta_{31})
(c_{23}^2-s_{23}^2)+(\tilde{A}_{\mu\tau})_k]}
{\tilde{\Delta}_{jk}^2\tilde{\Delta}_{ki}^2}
\cos \tilde{\Delta}_{ij}^{\prime}
\sin \tilde{\Delta}_{jk}^{\prime}
\sin \tilde{\Delta}_{ki}^{\prime},
\label{26} \\
\tilde{C}_{\mu\tau}&=&\sum_{(ij)}^{{\rm cyclic}}
\frac{-4[\Delta_{31}^2 s_{23}^2 c_{23}^2 c_{13}^4
(\lambda_i-a)(\lambda_j-a)
+(\tilde{C}_{\mu\tau})_{ij}]}
{\tilde{\Delta}_{ij} \tilde{\Delta}_{12}
\tilde{\Delta}_{23}\tilde{\Delta}_{31}}
\sin^2 \tilde{\Delta}_{ij}^{\prime}, \\
\tilde{D}&=&\sum_{(ijk)}^{{\rm cyclic}}\frac{-8 \Delta_{21}^2
(\lambda_k-a-\Delta_{31})^2
s_{12}^2 c_{12}^2 s_{23}^2 c_{23}^2 s_{13}^2}
{\tilde{\Delta}_{jk}^2\tilde{\Delta}_{ki}^2}
\cos \tilde{\Delta}_{ij}^{\prime}
\sin \tilde{\Delta}_{jk}^{\prime}
\sin \tilde{\Delta}_{ki}^{\prime} \label{27},
\end{eqnarray}
\end{widetext}
and the concrete expressions for $(\tilde{A}_{\mu\tau})_k$ and 
$(\tilde{C}_{\mu\tau})_{ij}$ are given in Appendix.
As for $\tilde{B}$, see (\ref{24}).
Note that these expressions (\ref{26})-(\ref{27}) are also exact. 
In the limit of small $\Delta_{21}$, 
$(\tilde{A}_{\mu\tau})_k$ and $(\tilde{C}_{e\tau})_{ij}$ are 
high order in $\Delta_{21}$ and can be neglected. 

Here, let us comment on the magnitude of each term.
$\tilde{D}$ is $O(\Delta_{21}^2 s_{13}^2)$ and 
hence it is difficult to observe 
the term proportional to $\cos 2\delta$.
$\tilde{A}_{\mu\tau}$ is suppressed by 
$\Delta_{21}s_{13}$ as in the previous cases.
In addition, $\tilde{A}_{\mu\tau}$ is also proportional 
to $c_{23}^2-s_{23}^2$.
Since the 2-3 mixing is almost maximal from the atmospheric 
neutrino experiments, $\tilde{A}_{\mu\tau}$ is considered 
to be rather small.
On the other hand, $\tilde{C}_{\mu\tau}$ 
has not any suppression factor.
Accordingly, the leading contribution comes from $\tilde{C}_{\mu\tau}$  
and the next to leading comes from $\tilde{B}$. 
Other contributions from $\tilde{A}_{\mu\tau}$ and $\tilde{D}$ 
are extremely small and we give 
the approximate formula for $\tilde{C}_{\mu\tau}$.

As in the previous cases, 
we obtain  
\begin{eqnarray}
\tilde{C}_{\mu\tau}\simeq 4s_{23}^2 c_{23}^2 c_{13}^4
\sin^2 \Delta_{31}^{\prime},
\end{eqnarray}
by neglecting the smallest effective mass 
$\lambda_1\simeq \Delta_{21}$.
We conclude that it is considerably difficult to observe the CP effects 
in the standard scenario.

\subsection*{\boldmath $P(\nu_e \to \nu_e)$, 
$P(\nu_{\mu} \to \nu_{\mu})$ and $P(\nu_{\tau} \to \nu_{\tau})$}

In this subsection, we calculate survival probabilities 
using unitarity and the transition probabilities 
derived in the previous subsections.

Let us first consider $P(\nu_e \to \nu_e)$.
The expression for the probability 
of this channel is needed, for example in neutrino 
factory experiments with the high energy 
$\nu_e (\bar{\nu}_e)$ beam.
In the standard scenario, the unitarity relation 
\begin{eqnarray}
P(\nu_e \to \nu_e)=1-P(\nu_e \to \nu_{\mu})
-P(\nu_e \to \nu_{\tau}) \label{28}
\end{eqnarray}
holds.
From (\ref{23}) and (\ref{24})
the coefficients of $\cos \delta$ and $\sin \delta$ 
are proportional to $J_r$ which includes $s_{23}c_{23}$.
As $P(\nu_e \to \nu_{\tau})$ is obtained 
from $P(\nu_e \to \nu_{\mu})$ by the exchange  
$s_{23} \to c_{23}, c_{23} \to -s_{23}$  
(i.e. $J_r \to -J_r$),
one can understand that the terms proportional to 
$\cos \delta$ and $\sin \delta$ disappear 
in (\ref{28}).
Therefore, $P(\nu_e \to \nu_e)$ is completely independent 
of $\delta$ and can be written in the form
\begin{eqnarray}
P(\nu_e \to \nu_e)=\tilde{C}_{ee},
\end{eqnarray}
as first pointed out in Ref. \cite{Kuo}.
Here, 
\begin{eqnarray}
\lefteqn{\tilde{C}_{ee}=1-\tilde{C}_{e\mu}-\tilde{C}_{e\tau}} \\
&=&1+\sum_{(ij)}^{{\rm cyclic}}
\frac{4[\Delta_{31}^2 s_{13}^2 c_{13}^2
\lambda_i \lambda_j
+(\tilde{C}_{ee})_{ij}]}
{\tilde{\Delta}_{ij} \tilde{\Delta}_{12}
\tilde{\Delta}_{23}\tilde{\Delta}_{31}}
\sin^2 \tilde{\Delta}_{ij}^{\prime}.
\end{eqnarray}
As for the concrete expression of 
$(\tilde{C}_{ee})_{ij}$, see Appendix.
Note that this expression is exact.
In the limit of small $\Delta_{21}$, 
$(\tilde{C}_{ee})_{ij}$ 
are higher order in $\Delta_{21}$ and can be ignored.
Furthermore, we obtain an approximate formula 
\begin{eqnarray}
\tilde{C}_{ee}\simeq 1-\frac{4\Delta_{31}^2 
s_{13}^2 c_{13}^2}
{(\Delta_{31}-a)^2}
\sin^2 (\Delta_{31}-a)^{\prime},
\end{eqnarray}
by neglecting the smallest effective mass 
$\lambda_1\simeq \Delta_{21}$.
$P(\nu_e \to \nu_e)$ mainly depends on 
$s_{13}$.  
This channel may be interesting 
because $s_{13}$ is determined regardless of   
the ambiguity of $\delta$.

Next, we calculate $P(\nu_{\mu} \to \nu_{\mu})$.
From unitarity, the relation  
$P(\nu_{\mu} \to \nu_{\mu})=1-P(\nu_{\mu} \to \nu_e)
-P(\nu_{\mu} \to \nu_{\tau})$
holds.
Then, the term proportional to $\sin \delta$ disappears 
and the probability can be written in the form 
\begin{eqnarray}
P(\nu_{\mu} \to \nu_{\mu})
=\tilde{A}_{\mu\mu}\cos \delta
+\tilde{C}_{\mu\mu}-\tilde{D}\cos 2\delta,
\end{eqnarray}
where the coefficients are given by 
\begin{eqnarray}
\tilde{A}_{\mu\mu}&=&-\tilde{A}_{e\mu}-\tilde{A}_{\mu\tau}
\simeq -\tilde{A}_{e\mu}, \label{29} \\
\tilde{C}_{\mu\mu}&=&1-\tilde{C}_{e\mu}-\tilde{C}_{\mu\tau}
\simeq 1-\tilde{C}_{\mu\tau} \label{30}.
\end{eqnarray}
In the approximation (\ref{29}) 
we use the fact that $\tilde{A}_{\mu\tau}$ 
is proportional to $c_{23}^2-s_{23}^2$ from (\ref{26})
and is very small in the case of maximal 2-3 mixing.
In the approximation (\ref{30}), we use the fact that 
$\tilde{C}_{e\mu}$ is suppressed by $s_{13}^2$. 
As for $\tilde{D}$, it is equivalent to that in 
$P(\nu_{\mu} \to \nu_{\tau})$. 

Finally, it may be useful to study the CP dependence of 
$P(\nu_{\tau} \to \nu_{\tau})$ although this channel 
is very difficult to observe.
From unitarity, the relation  
$P(\nu_{\tau} \to \nu_{\tau})=1-P(\nu_e \to \nu_{\tau})
-P(\nu_{\mu} \to \nu_{\tau})$ 
also holds.
Then, the probability can be written in the form 
\begin{eqnarray}
P(\nu_{\tau} \to \nu_{\tau})
=\tilde{A}_{\tau\tau}\cos \delta
+\tilde{C}_{\tau\tau}-\tilde{D}\cos 2\delta.
\end{eqnarray}
The coefficients are given by 
\begin{eqnarray}
\tilde{A}_{\tau\tau}&=&-\tilde{A}_{e\tau}-\tilde{A}_{\mu\tau}
\simeq -\tilde{A}_{e\tau}, \\
\tilde{C}_{\tau\tau}&=&1-\tilde{C}_{e\tau}-\tilde{C}_{\mu\tau}
\simeq 1-\tilde{C}_{\mu\tau}. 
\end{eqnarray}
Here, $\tilde{D}$ is also the same as that in  
$P(\nu_{\mu} \to \nu_{\tau})$.

\subsection*{Summary of CP dependences in Matter}

Here, we summarize the CP dependences of the oscillation 
probabilities for all channels in matter.
They are given by
\begin{eqnarray}
&&\hspace{-1cm}
P(\nu_e \to \nu_{\mu})=\tilde{A}_{e\mu}\cos\delta
+\tilde{B}\sin\delta+\tilde{C}_{e\mu}, \label{11} \\
&&\hspace{-1cm}
P(\nu_e \to \nu_{\tau})=-\tilde{A}_{e\mu}\cos\delta
-\tilde{B}\sin\delta+\tilde{C}_{e\tau}, \\
&&\hspace{-1cm}
P(\nu_{\mu} \to \nu_{\tau})=\tilde{A}_{\mu\tau}\cos\delta
+\tilde{B}\sin\delta+\tilde{C}_{\mu\tau}
+\tilde{D}\cos 2\delta, \\
&&\hspace{-1cm}
P(\nu_e \to \nu_e)=\tilde{C}_{ee}, \\
&&\hspace{-1cm}
P(\nu_{\mu} \to \nu_{\mu})=\tilde{A}_{\mu\mu}\cos\delta
+\tilde{C}_{\mu\mu}-\tilde{D}\cos 2\delta, \\
&&\hspace{-1cm}
P(\nu_{\tau} \to \nu_{\tau})=\tilde{A}_{\tau\tau}\cos\delta
+\tilde{C}_{\tau\tau}-\tilde{D}\cos 2\delta \label{12}. 
\end{eqnarray}
Comparing these results with (\ref{7})-(\ref{8}), 
we find that the probability for each channel in matter 
has the same form with respect to the CP phase $\delta$ as in vacuum.
It means that the matter effects just modify 
the coefficients.
Finally, we comment that the probabilities for T conjugate channels 
are obtained from (\ref{11})-(\ref{12}) by the replacement 
$\delta \to -\delta$ and 
those for CP conjugate channels are obtained by 
$\delta \to -\delta$ and $a \to -a$.

\section{Effective Mixing Angles and Effective CP phase}
\label{sec:5}

In this section, we show that our results (\ref{11})-(\ref{12}) 
with respect to the CP dependences are reproduced from 
the effective mixing angles and the effective 
CP phase calculated by Zaglauer and Schwarzer \cite{Zaglauer}. 
They have derived the effective mixing angles and CP phase 
from $|\tilde{U}_{\alpha\beta}|^2$ as 
\begin{widetext}
\begin{eqnarray}
\sin^2 \tilde{\theta}_{13}&=&
\frac{\lambda_3^2-\alpha \lambda_3+\beta}
{\tilde{\Delta}_{13}\tilde{\Delta}_{23}}, 
\label{31} \\
\sin^2 \tilde{\theta}_{12}
&=&\frac{-(\lambda_2^2-\alpha \lambda_2+\beta)\tilde{\Delta}_{31}}
{(\lambda_1^2-\alpha \lambda_1+\beta)\tilde{\Delta}_{32}
-(\lambda_2^2-\alpha \lambda_2+\beta)\tilde{\Delta}_{31}},
\\
\sin^2 \tilde{\theta}_{23} &=&
\frac{G^2 s_{23}^2+F^2 c_{23}^2+2GFs_{23}c_{23}\cos\delta}
{G^2+F^2}, \\
e^{-i\tilde{\delta}}&=&\frac{(G^2 e^{-i\delta}-F^2 e^{i\delta})
s_{23}c_{23}+GF(c_{23}^2-s_{23}^2)}
{\sqrt{(G^2 s_{23}^2+F^2 c_{23}^2+2GFs_{23}c_{23}\cos\delta)
(G^2 c_{23}^2+F^2 s_{23}^2-2GFs_{23}c_{23}\cos\delta)}}
\label{32},
\end{eqnarray}
\end{widetext}
where $\alpha$, $\beta$, $G$ and $F$ are given by  
\begin{eqnarray}
\alpha&=&p_{\mu\mu}+p_{\tau\tau}, \\
\beta&=&q_{ee}, \\
G s_{23}&=&p_{e\mu}^a \lambda_3+q_{e\mu}^a, \\
F c_{23}&=&p_{e\mu}^b \lambda_3+q_{e\mu}^b. 
\end{eqnarray}
Let us here describe the derivation of the effective 
mixing angles 
\footnote{Zaglauer and Schwarzer derive effective mixing 
angles by using $|\tilde{U}_{\alpha\beta}|^2$. 
In this paper, the expression for 
$\tilde{U}_{\alpha i}\tilde{U}_{\beta i}^*$ 
is used to avoid the complexity of calculation.}.
We can easily calculate the 1-3 mixing from 
$|\tilde{U}_{e3}|^2=\tilde{s}_{13}^2$. 
We can also calculate the 1-2 mixing   
by eliminating $\tilde{c}_{13}$ from 
$|\tilde{U}_{e1}|^2=\tilde{c}_{12}^2 
\tilde{c}_{13}^2$ and $|\tilde{U}_{e2}|^2=\tilde{s}_{12}^2 
\tilde{c}_{13}^2$.
Furthermore, the 2-3 mixing can be calculated by 
eliminating $\tilde{c}_{13}$ from the relations, 
\begin{equation}
|\tilde{U}_{\mu 3}|^2=\frac{|\tilde{U}_{e3}\tilde{U}_{\mu 3}^*|^2}
{|\tilde{U}_{e3}|^2}
=\frac{|p_{e\mu}\lambda_3+q_{e\mu}|^2}
{\tilde{\Delta}_{31}\tilde{\Delta}_{32}
(\lambda_3^2-\alpha\lambda_3+\beta)}
=\tilde{s}_{23}^2 \tilde{c}_{13}^2, 
\end{equation}
\begin{equation}
|\tilde{U}_{\tau 3}|^2=\frac{|\tilde{U}_{e3}\tilde{U}_{\tau 3}^*|^2}
{|\tilde{U}_{e3}|^2}
=\frac{|p_{e\tau}\lambda_3+q_{e\tau}|^2}
{\tilde{\Delta}_{31}\tilde{\Delta}_{32}
(\lambda_3^2-\alpha\lambda_3+\beta)}=\tilde{c}_{23}^2 
\tilde{c}_{13}^2.
\end{equation}
Next, we consider the CP dependence.
One may think that the CP dependence of (\ref{32}) 
is very complicated due to $\cos \delta$ in the denominator 
at a glance.
However, our results (\ref{11})-(\ref{12}) 
are very simple and the probabilities 
are linear in $\sin \delta$, $\cos \delta$ and 
$\cos 2\delta$.
It suggests that the complicated CP dependence 
in $e^{-i\tilde{\delta}}$ cancels 
with that in other terms.
From (\ref{31})-(\ref{32}), $\delta$ is only included in 
$\sin \tilde{\theta}_{23}$ except for $e^{-i\tilde{\delta}}$.
Furthermore, the numerator of $\sin \tilde{\theta}_{23}$ 
is partially in agreement with the denominator 
of $e^{-i\tilde{\delta}}$.
The product of $\tilde{s}_{23}$, $\tilde{c}_{23}$ and 
$e^{-i\tilde{\delta}}$ is calculated as 
\begin{equation}
\tilde{s}_{23} \tilde{c}_{23} e^{-i\tilde{\delta}}
=\frac{(G^2 e^{-i\delta}-F^2 e^{i\delta})
s_{23}c_{23}+GF(c_{23}^2-s_{23}^2)}{G^2+F^2} \label{33}.
\end{equation}
Note that the CP phase in the denominator of $e^{-i\tilde{\delta}}$
completely disappears.
The imaginary part of (\ref{33}) is known as 
the Toshev identity \cite{Toshev} 
\begin{equation}
\tilde{s}_{23} \tilde{c}_{23} \sin \tilde{\delta}
= \tilde{s}_{23} \tilde{c}_{23} \sin \tilde{\delta},
\label{33-1}
\end{equation}
which is independent of matter effects. 

Finally, we demonstrate how the CP dependences of 
the oscillation probabilities become simple 
taking the channel $P(\nu_e \to \nu_{\mu})$ 
as an example.
In general, the oscillation probabilities in matter 
are also obtained from those in vacuum by the exchange  
$\theta_{ij} \to \tilde{\theta}_{ij}$ and 
$\delta \to \tilde{\delta}$. 
Namely, from (\ref{4})-(\ref{5}) and (\ref{6}), 
$P(\nu_e \to \nu_{\mu})$ in matter is constituted by 
\begin{eqnarray}
{\rm Re}\tilde{J}_{e\mu}^{12}
&=&
-(\tilde{c}_{12}^2-\tilde{s}_{12}^2)\tilde{J_r} 
\cos\tilde{\delta} \nonumber \\
&&+\tilde{s}_{12}^2 \tilde{c}_{12}^2 \tilde{c}_{13}^2
(\tilde{s}_{23}^2 \tilde{s}_{13}^2-\tilde{c}_{23}^2), 
\\
{\rm Re}\tilde{J}_{e\mu}^{23}&=&\tilde{J_r} \cos\tilde{\delta}
-\tilde{s}_{12}^2 \tilde{s}_{23}^2 \tilde{s}_{13}^2 
\tilde{c}_{13}^2, \\
{\rm Re}\tilde{J}_{e\mu}^{31}&=&-\tilde{J_r} \cos\tilde{\delta}
-\tilde{c}_{12}^2 \tilde{s}_{23}^2 \tilde{s}_{13}^2 \tilde{c}_{13}^2, \\
\tilde{J}&=&\tilde{J_r} \sin \tilde{\delta}.
\end{eqnarray}
We can see that $\sin\tilde{\delta}$ ($\cos\tilde{\delta}$) 
always appears together with $\tilde{J}_r$ in 
${\rm Re}\tilde{J}_{e\mu}^{ij}$ or in $\tilde{J}$.
On the other hand, the CP dependence for the product 
$\tilde{s}_{23}\tilde{c}_{23}e^{-i\tilde{\delta}}$ 
become simple as shown in (\ref{33}). 
Therefore, the CP dependence of the probability is simplified 
although that of the effective CP phase is complicated. 
Then, we obtain the same results 
\begin{eqnarray}
P(\nu_e \to \nu_{\mu})=\tilde{A}_{e\mu}\cos \delta
+\tilde{B}\sin \delta +\tilde{C}_{e\mu}
\end{eqnarray}
as (\ref{21}). 
As for other channels, we obtain the same results 
as those calculated in our approach.
This is guaranteed by the Toshev identity.

\section{New Matter Invariant Identity}
\label{sec:6}

In the previous section, we rederived two matter invariant identities.
One is the Naumov-Harisson-Scott identity (\ref{18-1})
and second is the Toshev identity (\ref{33-1}).
We obtain a new identity
\begin{equation}
\tilde{\Delta}_{12}\tilde{\Delta}_{23}\tilde{\Delta}_{31}
\tilde{s}_{12}\tilde{c}_{12}\tilde{s}_{13}\tilde{c}_{13}^2
=\Delta_{12}\Delta_{23}\Delta_{31}s_{12}c_{12}s_{13}c_{13}^2
\label{21-1}
\end{equation}
related to the 1-2 mixing and the 1-3 mixing by dividing 
the Naumov-Harrison-Scott identity by the Toshev identity.

Here, we give another proof of this new identity 
(\ref{21-1}).
We can independently derive the identity (\ref{21-1}) without 
using other identities in this proof.
At first, taking $k_1, k_2$ as the eigenvalues 
of sub-matrix 
\begin{eqnarray}
\frac{1}{2E}\left(
\begin{array}{cc}
H_{\mu\mu} & H_{\mu\tau} \\
H_{\tau\mu} & H_{\tau\tau}
\end{array}
\right),
\end{eqnarray}
we rewrite $|\tilde{U}_{ei}|^2$ obtained from (\ref{16}) as 
\begin{eqnarray}
|\tilde{U}_{ei}|^2&=&\frac{\lambda_i^2-(p_{\mu\mu}+p_{\tau\tau})
\lambda_i+q_{ee}}
{\tilde{\Delta}_{ji}\tilde{\Delta}_{ki}} \nonumber \\
&=&\frac{(\lambda_i-k_1)(\lambda_i-k_2)}
{\tilde{\Delta}_{ji}\tilde{\Delta}_{ki}}. 
\end{eqnarray}
Then, we obtain 
\begin{eqnarray}
\lefteqn{(\tilde{\Delta}_{12}\tilde{\Delta}_{23}\tilde{\Delta}_{31})^2
|\tilde{U}_{e1}|^2 |\tilde{U}_{e2}|^2 |\tilde{U}_{e3}|^2} 
\nonumber \\
&&\hspace{-0.5cm}=\prod_{i=1,2}(\lambda_1-k_i)(\lambda_2-k_i)(\lambda_3-k_i) 
\nonumber \\
&&\hspace{-0.5cm}=\prod_{i=1,2}(m_1^2-k_i)(m_2^2-k_i)(m_3^2-k_i)={\rm const}.
\label{34},
\end{eqnarray}
where we use the relations obtained from 
the characteristic equation of Hamiltonian 
\begin{widetext} 
\begin{eqnarray}
\lambda_1+\lambda_2+\lambda_3&=&m_1^2+m_2^2+m_3^2+a, \\
\lambda_1 \lambda_2+\lambda_2 \lambda_3+\lambda_3 \lambda_1
&=&m_1^2 m_2^2+m_2^2 m_3^2+m_3^2 m_1^2+a(k_1+k_2), \\
\lambda_1 \lambda_2 \lambda_3&=&m_1^2 m_2^2 m_3^2+ak_1 k_2,
\end{eqnarray}
\end{widetext}
from the first line to the second line in (\ref{34}).
Note that $k_i$ is the eigenvalue of the sub-matrix 
of Hamiltonian in vacuum and then does not depend on 
matter effects.
Therefore, (\ref{34}) is a matter invariant quantity and 
it is rewritten as (\ref{21-1}). 

\section{Summary}
\label{sec:7}

We summarize the results obtained in this paper.
We consider the probabilities and the 
CP dependences for all channels within the context 
of standard three neutrino oscillations in constant matter.

\begin{enumerate}
\renewcommand{\labelenumi}{(\roman{enumi})}

\item  We have derived an exact and simple formula 
for the oscillation probability applicable to all channels.
We have also investigated the CP dependences of various probabilities 
using this formula.
As the results, $P(\nu_e \to \nu_{\mu})$ can be written in the form 
$A_{e\mu}\cos\delta+B\sin\delta+C_{e\mu}$.
$P(\nu_e \to \nu_{\tau})$ has the same 
form as $P(\nu_e \to \nu_{\mu})$.
On the other hand, the probability for $\nu_{\mu}$ to $\nu_{\tau}$ 
transition can be written as $P(\nu_{\mu} \to \nu_{\tau})=
A_{\mu\tau}\cos\delta+B\sin\delta+C_{\mu\tau}+D\cos 2\delta$.
In survival probabilities $P(\nu_e \to \nu_e)$, 
$P(\nu_{\mu} \to \nu_{\mu})$ and $P(\nu_{\tau} \to \nu_{\tau})$, 
there is no $\sin\delta$ term.
As for $P(\nu_e \to \nu_e)$, $\cos\delta$ term also 
disappears and the probability is completely independent 
of $\delta$.
We have found that the probability for each channel in matter 
has the same form with respect to the CP phase as in vacuum.
That is, the matter effects just modify 
the coefficients $A$, $B$, $C$ and $D$. 
We have also given the exact expression for the coefficients 
in constant matter.

\item  We have shown that our results with respect to the CP dependences 
are reproduced from the effective mixing angles 
and CP phase derived by Zaglauer and Schwarzer.
Although the effective CP phase are complicated, 
the CP dependence of probability is simplified for all channels. 
This is guaranteed by Toshev identity.
Finally we have obtained a new identity 
related to the 1-2 mixing and the 1-3 mixing
by dividing the Naumov-Harrison-Scott identity
by the Toshev identity.
\end{enumerate}

\begin{acknowledgments}

The authors would like to thank Prof. A. I. Sanda  
for making a number of 
helpful suggestions.
We would like to thank Prof. H. Minakata and Prof. O. Yasuda 
for discussions and valuable comments.

\end{acknowledgments}

\appendix

\section{Subleading Terms in the $\tilde{A}_{\alpha\beta}$,
$\tilde{C}_{\alpha\beta}$}

Here, we describe the terms which include higher order of 
$\Delta_{21}$ explicitly in 
$\tilde{A}_{\alpha\beta}$ and $\tilde{C}_{\alpha\beta}$.

First, the terms in $P(\nu_e \to \nu_{\mu})$ are 
\begin{widetext}
\begin{eqnarray}
(\tilde{A}_{e\mu})_k&=&\Delta_{21}^2 J_r\times 
\left[\Delta_{31}\lambda_k
(c_{12}^2-s_{12}^2)+\lambda_k^2 s_{12}^2
-\Delta_{31}^2 c_{12}^2 \right],  \\
(\tilde{C}_{e\mu})_{ij}
&=&\Delta_{21}s_{13}^2 \times \left[\Delta_{31}\{
-\lambda_i(\lambda_j s_{12}^2+\Delta_{31}c_{12}^2)
-\lambda_j(\lambda_i s_{12}^2+\Delta_{31}c_{12}^2)\}
s_{23}^2 c_{13}^2 \right] \nonumber \\
&+&\Delta_{21}^2 \times
\left[(\lambda_i-\Delta_{31})(\lambda_j-\Delta_{31})
s_{12}^2 c_{12}^2 c_{23}^2 c_{13}^2 \right] 
\nonumber \\
&+&\Delta_{21}^2 s_{13}^2 \times
\left[(\lambda_i s_{12}^2+\Delta_{31}c_{12}^2)
(\lambda_j s_{12}^2+\Delta_{31}c_{12}^2)
s_{23}^2 c_{13}^2 \right], 
\end{eqnarray}
where the second line of $(\tilde{C}_{e\mu})_{ij}$ is relatively large 
in the case of $\Delta_{21}/\Delta_{31}< s_{13}$ 
so that there is no suppression factor $s_{13}^2$.

Second, the terms related to $P(\nu_e \to \nu_{\tau})$ are 
\begin{eqnarray}
(\tilde{C}_{e\tau})_{ij}
&=&
\Delta_{21}s_{13}^2 \times \left[\Delta_{31}\{
-\lambda_i(\lambda_j s_{12}^2+\Delta_{31}c_{12}^2)
-\lambda_j(\lambda_i s_{12}^2+\Delta_{31}c_{12}^2)\}
c_{23}^2 c_{13}^2 \right] 
\nonumber \\
&+&\Delta_{21}^2 \times 
\left[(\lambda_i-\Delta_{31})(\lambda_j-\Delta_{31})
s_{12}^2 c_{12}^2 s_{23}^2 c_{13}^2 \right] 
\nonumber \\
&+&\Delta_{21}^2 s_{13}^2 \times
\left[(\lambda_i s_{12}^2+\Delta_{31}c_{12}^2)
(\lambda_j s_{12}^2+\Delta_{31}c_{12}^2)
c_{23}^2 c_{13}^2 \right],
\end{eqnarray}
where there is also no suppression factor $s_{13}^2$ 
in the second line.

Third, the terms related to $P(\nu_{\mu} \to \nu_{\tau})$ are 
\begin{eqnarray}
(\tilde{A}_{\mu\tau})_k&=&
-\Delta_{21}^2 s_{13}(c_{23}^2-s_{23}^2) \nonumber \\
&&\times 
\left[s_{12}c_{12}s_{23}c_{23}
(\lambda_k-a-\Delta_{31})\{\Delta_{31}(s_{12}^2-s_{13}^2 c_{12}^2)
+(\lambda_k-a-\Delta_{31})c_{12}^2\}\right] \\
(\tilde{C}_{\mu\tau})_{ij}
&=&\Delta_{21}\times \left[\Delta_{31}\{
\Delta_{31}(\lambda_i+\lambda_j-2a)(c_{13}^2+c_{12}^2)
-2(\lambda_i-a)(\lambda_j-a)c_{12}^2\}
s_{23}^2 c_{23}^2 c_{13}^2 \right] \nonumber \\
&+&\Delta_{21}s_{13}^2 \times 
\left[\Delta_{31}\{-\Delta_{31}(\lambda_i+\lambda_j-2a)
+2(\lambda_i-a)(\lambda_j-a)\}s_{12}^2 s_{23}^2
c_{23}^2 c_{13}^2\right] \nonumber \\
&+&\Delta_{21}^2 \times \left[\{
\Delta_{31}c_{13}^2+(\lambda_i-a-\Delta_{31})c_{12}^2\} 
\{\Delta_{31}c_{13}^2+(\lambda_j-a-\Delta_{31})c_{12}^2\}
s_{23}^2 c_{23}^2\right] \nonumber \\
&+&\Delta_{21}^2 s_{13}^2\times \left[-\Delta_{31}
(\lambda_i+\lambda_j-2a-2\Delta_{31})s_{12}^2 s_{23}^2 c_{23}^2 c_{13}^2
\right. \nonumber \\ 
&&\left.\hspace{0.5cm}
+(\lambda_i-a-\Delta_{31})(\lambda_j-a-\Delta_{31})
\{s_{12}^2 c_{12}^2 (c_{23}^2-s_{23}^2)
+s_{12}^4 s_{23}^2 c_{23}^2 s_{13}^2\}\right].
\end{eqnarray}
Finally, 
the terms related to $P(\nu_e \to \nu_e)$ are 
\begin{eqnarray}
(\tilde{C}_{ee})_{ij}
&=&\Delta_{21}s_{13}^2 \times \left[\Delta_{31}\{
-\lambda_i(\lambda_j s_{12}^2+\Delta_{31}c_{12}^2)
-\lambda_j(\lambda_i s_{12}^2+\Delta_{31}c_{12}^2)\}
c_{13}^2 \right]
\nonumber \\
&+&\Delta_{21}^2 \times
\left[(\lambda_i-\Delta_{31})(\lambda_j-\Delta_{31})
s_{12}^2 c_{12}^2 c_{13}^2 \right] \nonumber \\
&+&\Delta_{21}^2 s_{13}^2 \times 
\left[(\lambda_i s_{12}^2+\Delta_{31}c_{12}^2)
(\lambda_j s_{12}^2+\Delta_{31}c_{12}^2)
c_{13}^2 \right], 
\end{eqnarray}
\end{widetext}
where there is no suppression factor $s_{13}^2$ 
in the second line as in $P(\nu_e \to \nu_{\mu})$ or 
$P(\nu_e \to \nu_{\tau})$. 
Therefore, the contribution is relatively large 
in the case of $\Delta_{21}/\Delta_{31}<s_{13}$.

\end{document}